\documentclass[12pt,a4paper]{article}

\usepackage[a4paper, top=2.5cm, bottom=2.5cm, left=2.5cm, right=2.5cm]{geometry}

\usepackage{titling}
\usepackage[utf8]{inputenc}
\usepackage[T1]{fontenc}
\usepackage{mathptmx}

\usepackage{amsmath, amssymb, amsthm}

\usepackage{graphicx}
\usepackage{float}
\usepackage{booktabs}
\usepackage{multirow}
\usepackage{longtable}
\usepackage{tabularx}
\usepackage{caption}
\usepackage{subcaption}

\usepackage{tikz}
\usetikzlibrary{shapes.geometric, arrows, positioning, fit}

\usepackage[numbers]{natbib}
\usepackage[colorlinks=true, linkcolor=blue, citecolor=blue, urlcolor=blue]{hyperref}

\tikzstyle{block} = [rectangle, draw, minimum width=4.5cm, minimum height=1cm, text centered]
\tikzstyle{arrow} = [thick,->,>=stealth]
\tikzstyle{header} = [rectangle, draw, fill=yellow!80, minimum width=10cm, minimum height=1cm, text centered]
\tikzstyle{stage} = [rectangle, draw, fill=blue!20, minimum width=1.5cm, minimum height=2cm, text centered, rotate=90]

\begin{document}

\title{Entropy-Based Methods to Address Sampling Bias in Archaeological Predictive Modeling}

\author{
    \begin{tabular}[t]{@{}c@{}}
        Mehmet Sıddık Çadırcı\textsuperscript{1}, \quad
        Golnaz Shahtahmassebi\textsuperscript{2}\thanks{Corresponding author: \href{mailto:golnaz.shahtahmassebi@ntu.ac.uk}{golnaz.shahtahmassebi@ntu.ac.uk}}
    \end{tabular}
}

\date{
    \small
    \textsuperscript{1}Faculty of Science, Department of Statistics, Cumhuriyet University, Sivas, Turkey \\
    \textsuperscript{2}Department of Computer Science, Nottingham Trent University, Nottingham NG11 8NS, UK
}

\maketitle

\begin{abstract}

Predictive modeling in archaeology is essential for the understanding of people's behavior in the past and for guiding heritage conservation. However, spatial sampling bias caused by uneven research effort can severely limit model reliability. This research describes a novel new framework that integrates entropy-based corrections to measure and minimize such biases in archaeological modeling of foresight. Leveraging the open access data of the Grand Staircase-Escalante National Monument, we employ Shannon entropy to determine survey coverage and assign appropriate weights to pseudo-absence points. We combine these weights with predictive models such as Bayesian Spatial Logistic Regression (via R-INLA), Generalized Additive Models, Maximum Entropy and Random Forests. Our findings prove that entropy-aware models exhibit improved accuracy and robustness, especially for under-surveyed regions. This approach not only advances methodological transparency, but also improves the interpretation of archaeological prediction under conditions of data uncertainty. The proposed framework offers a scalable, theoretically grounded strategy for addressing spatial bias in archaeological datasets.

\end{abstract}

\section{INTRODUCTION}\label{sec:intro} 

Predictive modeling to predict where archaeological sites may be depending on environmental and spatial variables has become an increasingly indispensable tool in archaeological research and cultural heritage management. It supports the prioritization of research activities, provides guidance for identifying mitigation strategies, and illustrates comprehensive patterns of past human activity across large and various landscapes. Planning at a regional scale or academic research are activities in which today's archaeologists employ predictive modeling of the location of sites in an increasingly predictive approach. Today, these models essentially become tools for balancing heritage conservation and economic development, given increasing development threats and limited resources for fieldwork.

This unpredictability in archaeological predictive models arises from spatial sampling bias. However, unlike ecological or remote sensing surveys, which rely on a defined sampling design, archaeological surveys are normally determined by practical considerations - accessibility of land, subdivision of the landscape into plots, construction of infrastructure, etc. This creates the result of uneven coverage: Some sites are fully explored, while others are virtually undisturbed. Operating on biased observations for model training means that models tend to sell on over-explored areas and underestimate the presence of sites in remote or under-studied areas. Bias, in another form, tends to create a more damaging feedback cycle: According to the model, certain researched fields may be represented as high-probability sites (because they have yielded the most known sites), resulting in more research in the same areas, while neglecting hard-to-access areas. Although researchers have been recognizing these problems for decades, addressing survey bias has largely been done in a haphazard ways (through spatial filtering of data, weighting of background samples and the like) and without integrating such methods into the traditional practice, suggesting that a more principled approach is needed.

Compounding these issues, many archaeological predictive models rely on presence/pseudo-absence data rather than true absence data. In practice, genuine absence points are rarely available in archaeology \citep{yaworsky2020advancing}, so researchers generate pseudo-absence points by assuming no sites exist in unsurveyed locations. These pseudo-absences can inadvertently mirror survey gaps rather than actual site absences, further skewing model outcomes by essentially marking poorly surveyed areas as “site-free.” For example, an algorithm may erroneously conclude that an uninvestigated valley contains no archaeological sites simply because none have been recorded there, engendering a false sense of certainty about an area never properly examined. The persistent combination of sampling bias and uncertain absences has prompted calls for predictive frameworks that explicitly account for survey effort and spatial uncertainty, ensuring that model predictions are not simply reproducing survey imbalances.

In this paper, we introduce a novel methodology in the field of archaeological predictive modeling that combines random sampling concepts with statistical entropy \citep{shannon1948mathematical} and is applied as a diagnostic and correction tools for sampling bias. The entropy measure is based on information theory and provides a quantitative uncertainty about the spatial distribution of the research effort. By examining the entropy of research on the study area, we can determine our under-sampled regions and characterize where our knowledge of site distribution is most precarious. Alternatively, these entropy values are included in the model-training process by weighting the pseudo-absence data according to their estimated reliability - training points in low-entropy (well-surveyed) regions are implicitly favored, while those in high-entropy (poorly-surveyed) locations receive no preference. This mechanism of weighting is a proper way to decrease the weighting of data coming from high uncertainty areas, thus compensating for the uneven research effort during the modeling.

There is no doubt that the entropy approach works effectively on this large public dataset from the Grand Staircase-Escalante National Monument in Utah. It is a particularly abundant dataset previously used for testing between competing Varietal models \citep{yaworsky2020advancing}, enabling us to assess how taking survey bias accounting can improve forecasting performance. For analysis, we consider several popular forecasting models, including Bayesian spatial logistic regression, generalized additive models (GAMs), maximum entropy (MaxEnt), and random forests (RF), and we weight the training data with an entropy measure in some versions and leave the data unweighted in others. These experiments will provide insights into the extent to which more accurate predictions and robustness resulting from deploying entropy-based weights may be available, especially in unexplored or under-explored areas and thus a source of complications for traditional modeling.

This study aims to improve the interpret-ability, generalisability and fairness of archaeological site predictions by combining entropy analysis with state-of-the-art predictive modeling techniques. The proposed framework therefore strengthens the methodological rigour of spatial modeling in archaeology, providing a replicable approach to solving sampling bias issues in other disciplines that depend on incomplete or opportunistic data collection, such as ecology and paleontology.  The structure of the paper is as follows: Section ~\ref{sec:lit_review} discusses relevant contributions to archaeological predictive modelling, with a particular focus on presence-only modelling, entropy-based approaches and spatial statistics. Section ~\ref{sec:meth} presents the methodology, including data pre-processing steps, the construction of entropy-based weights, and the application of four modelling frameworks: Bayesian spatial logistic regression, generalised additive models (GAMs), MaxEnt and Random Forests. The results and analyses from the comparative experiments are presented in Section ~\ref{sec:res_dis} with particular attention paid to the effect of entropy correction on predictive accuracy, generalisation and overall model behaviour. The wider methodological and practical implications of entropic bias correction in archaeological modelling workflows are then reflected on in Section ~\ref{sec:conc}.

\section{LITERATURE REVIEW}\label{sec:lit_review}
The application of Maximum Entropy (MaxEnt) modeling and statistical entropy has become prominent in archaeology and ecology over recent years. The literature review synthesizes the contemporary uses of MaxEnt and statistical entropy, stressing their strengths and weaknesses, including sampling bias and data non-representability. MaxEnt has been the go-to solution for predictive models in archaeology and ecology, mainly because of its ability to process presence-only data. Thus, Rafuse showed the use of MaxEnt to predict hunter-gatherer sites in Southern Pampas, Argentina, despite problems related to non-representative sampling and environmental conditions (Rafuse, 2021) \citep{rafuse2021maxent}. In the same vein, Wang et al. showed that MaxEnt provides good predictions for archaeological sites in Japan and China, adding to the idea that these sites are not randomly distributed but instead somehow related to environmental characteristics (Wang, 2023) \citep{wang2023archaeological}. This is supported by the study of Mcmichael et al., who observed that MaxEnt models are those most commonly used to predict species distributions in ecological studies, thereby indicating that the model is highly relevant in studies across disciplines (McMichael et al., 2017) \citep{mcmichael2017ancient}. MaxEnt has therefore been used effectively to model archaeological sites, ancient human activity, and settlement. For example, Howey et al. used MaxEnt to investigate monument construction over time in Michigan, enabling it to address complex societal developments (Howey et al., 2016) \citep{howey2016geospatial}. Furthermore, Yaworsky et al. evaluated various statistical approaches, including MaxEnt, to predict archaeological site locations, thus underscoring the importance of rigorous statistics in archaeology (Yaworsky et al., 2020) \citep{yaworsky2020advancing}. However, such models suffer in practice due to data limitations, mainly when training datasets are small or biased towards certain domains (Yaworsky et al., 2020). Despite its advantages, MaxEnt and similar types of modeling raise issues related to sampling bias and representativeness of the data. Research has demonstrated that forecasting models based on presence data can be heavily affected by sampling bias and spatial autocorrelation, and can provide inaccurate forecasts (Souza et al., 2018) \citep{de2018pre}. For example, Guedes et al. ephasized the need for spatial filtering to counteract the effects of biased occurrence datasets in archaeological modeling, since they suspected traditional background point manipulation in MaxEnt may not fully resolve that (Souza et al., 2018)\citep{de2018pre}. Likewise, Cano et al. posited that to ensure greater precision, ecological niche modeling with MaxEnt ought to factor in data constraints (Cano et al., 2023)\citep{ortiz2023ecological}. Yet, entropy-related techniques in ecology have further complicated the matter. Although this maximum entropy principle has been put to use for analyzing species distribution and abundance, the utility of the principle depends on the nature of the data sets available-favorability and completeness (Favretti, 2017; Haegeman $\And$ Etienne, 2010)  \citep{favretti2017remarks, haegeman2010entropy}. Upon examination, according to Favretti, the Maximum Entropy Principle tends to be used to infer distributions from macroscopic information, and its use is complicated when the dataset is incomplete (Favretti, 2017) \citep{favretti2017remarks}. This issue is even more relevant within archaeological settings, where the archaeological data are sometimes scarce and unevenly distributed across landscapes. In conclusion, while MaxEnt and statistical entropy remain valuable tools, entropy theories in archaeology and ecology need to address significant issues such as sampling bias and unrepresentative data. Furthermore, future research should seek to develop methodologies that improve both the quality and representativeness of a dataset to improve the reliability of predictive models. This could include being able to combine different data sources, utilizing diverse and sophisticated statistical techniques, and testing models very well to discover whether they are truly representative of historical and ecological realities.

\textbf{MaxEnt and Presence-Only Predictive Modeling} The MaxEnt method has been a distinctive tool in archaeological and ecological cases, given that it allowed interaction with presence-only data. Likewise, Rafuse (2021) \citep{rafuse2021maxent} illustrates the power of MaxEnt modeling to identify hunter-gatherer locations in Argentina, even when facing problems of environmental bias and heterogeneous sampling. Wang et al. (2023) \citep{wang2023archaeological} also employed MaxEnt in East Asia to validate the idea that archaeological sites correspond to specific environmental characteristics and thus are not randomly distributed. Similarly, McMichael et al. (2017) and Howey et al. (2016) \citep{mcmichael2017ancient, howey2016geospatial} used MaxEnt to investigate ancient land use and monument construction in ecological and cultural settings. Yaworsky et al. (2020) \citep{yaworsky2020advancing} represents a comparative field modeling of MaxEnt and other statistical tools and highlights the value of machine learning techniques in field forecasting.

\textbf{Sampling Bias and Spatial Autocorrelation.} MaxEnt is exclusively an asset-based data-driven modeling tool; hence, it is subject to spatial sampling bias. Souza et al. (2018)\citep{de2018pre} emphasize that survey bias and autocorrelation can lead to distortion of model outputs and propose spatial filtering and bias correction methods to mitigate it. Similarly, Cano et al. (2023) highlight that \citep{ortiz2023ecological} It cautioned that MaxEnt is optimal only when the occurrence data is representative. Studies such as Fourcade et al. (2014) \citep{Fourcade2014} address spatial sampling bias in asset-only forecasting modeling (MaxEnt). They have consistently demonstrated that systematic spatial sampling (filtering) of occurrence records is among the best methods to reduce geographic sampling bias. These insights highlight the continuing need for models that incorporate spatial structure and research effort directly into the modeling process.

\textbf{Entropy-Based Modeling and Information Theory. } The term “entropy” has recently been used to describe the uncertainty or variability associated with data distributions. To mention two examples from entropy, the Maximum Entropy Principle describes the abundance and spatial distribution of species \citep{favretti2017remarks, haegeman2010entropy}. These authors show that entropy-based modeling strongly depends on the quality of input data and equilibrium assumptions. At the same time, entropies remain an underutilized tool in archaeology; here, a test of the use of entropy to measure the inequality of research coverage.

\textbf{Developments in Bayesian and Spatial Statistical Methods} To address the limitations of classical approaches, modern analysis has turned to the Bayesian spatial paradigm, which explicitly addresses spatial autocorrelation and uncertainty. Gelfand and Banerjee (2003) \citep{banerjee2003hierarchical} It addresses the utility of hierarchical spatial models in geographic applications, including CAR priorities. Bakka et al. (2018) \citep{bakka2018spatial} describe how the R-INLA framework can be leveraged to perform fast Bayesian inference for spatial GLMs, including logistic regression with structured priors. These approaches are particularly well adapted to archaeological situations, given that research data is often highly spatially structured and heterogeneous. While MaxEnt, statistical entropy, and some spatial techniques have each targeted components of the modeling and prediction problem, studies that present a more systematic integration of entropy-based bias correction into machine learning and Bayesian spatial frameworks are uncommon. This paper accomplishes this integration by evaluating entropy-weighted models under a wide range of statistical paradigms and testing the membership of such models under known conditions of sample selection bias.

\section{METHODOLOGY}\label{sec:meth}

\subsection*{Data Collection and Preprocessing}
We compiled a dataset containing a list of known archaeological sites from the study area and environmental determinants. Site occurrence data were collected from systematic surveys and historical sources and each site was assigned a geographic coordinate. Landform configuration, hydrology and other landscape factors served as environmental covariates in the study; indices of elevation, slope, distance to water surfaces, soil type and vegetation cover were included. These were retrieved from GIS databases and resampled at a common spatial resolution. The data was then pre-processed to prepare it for modeling. Pre-processing of field coordinates involved cleaning and georeferencing them, while environmental variables involved processing missing and outliers and standardizing continuous predictors to comparable scales.

Pseudo-absences (background points) were created for models that require negative samples to validate the method without true absence data (locations known to be investigated with no findings). Background points were randomly sampled over the entire field of study, although the sampling probability was adjusted according to the research effort extracted to avoid introducing further bias (see below). Thus, these pseudo-absence points are more likely to be selected in well-surveyed areas and less likely in un-surveyed areas, which aligns with the detection process. Such pseudo-absence points were assigned a response value of 0 (absence) and combined with existing ones (value 1) to train logistic regressions, GAMs and random forest models.

\subsection*{Entropy Based Bias Measurement and Weighting}
Archaeological and survey data often suffer from spatial biases, with some areas being the subject of more research than others. We characterized the unbiased survey effort through an entropy-based measure of sampling distribution. The study region was divided into spatial units (which could be grid cells or administrative districts) to assess survey coverage. For each $i$ unit, we calculate the proportion $p_i$ of all known sites (or total survey observations) located in that unit. A measure of how dispersed the survey work is is summarized by the Shannon entropy of the distribution $P=\{p_i\}$ given by
\[ H(P) = -\sum_{i} p_i \ln p_i, \] 
low entropy means that only a few locations dominate the survey data (high bias) and higher entropy with upper bound $\ln N$ for $N$ units means that they are surveyed more or less equally (low bias) \citep{shannon1948mathematical}. In this way, by applying a measure of bias, weights were derived to correct the sampling effort estimate during modeling. Each presence record was assigned a weight inversely proportional to the probability of the survey taking place in that region. The practical effect is, therefore, as follows: $p_i$ sites in regions with a high contribution to $p_i$ were assigned smaller weights, whereas sites in regions with a low level of surveys were assigned larger weights. Weights were therefore used in the final model to balance the effect of training samples: logistic regression, GAM and random forest were equipped with sample weights in the presence and pseudo-absence cases. At the same time, a bias grid (sampling probability surface) was provided to the MaxEnt Model so that background points were placed according to the survey effort. By using these entropy-based weights, we reduced model bias due to survey density, giving us predictions more representative of site fidelity.

\subsection*{Predictive Modeling Approaches} 
Predictive modeling was attempted through four different methods applied to predict the occurrence of archaeological sites, as discussed in detail below. Each method provides a different trade-off in interpretability and flexibility, and our results should prove robust to all methods.

\subsection*{Perform Bayesian Spatial Logistic Regression with CAR Prior}
For the analysis, we employed a Bayesian spatial logistic regression model with CAR before discussing spatial autocorrelation of sites. The model assumes the log-odds of site availability at location $i$ as a linear estimator with a random effect specific to each site: 
\[ \text{logit}\,\big(P(Y_i=1)\big) = \beta_0 + \sum_{j=1}^{p} \beta_j\,x_{ij} + \phi_i, \]  
where $Y_i$ is the presence indicator, $x_{ij}$ is the value of predictor $j$ for location $i$, and $\phi_i$ is the spatial random effect term. The random effects are given an intrinsic CAR prior $\{\phi_i\}$, which creates spatial smoothing by encouraging neighboring locations to have similar $\phi_i$ values \citep{besag1991bayesian}. The authors therefore adapt this hierarchical model from a Bayesian perspective (using Markov Chain Monte Carlo sampling) to obtain posterior distributions of regression coefficients and spatial random effects. Applying the CAR prior absorbs the remaining spatial structure from the data and thus improves the prediction performance when sites are spatially clustered.

\subsection*{Generalized Additive Models (GAMs)}
GAM capture the nonlinear relationships between environmental variables and field presence \citep{hastie1986generalized}. A GAM extends a generalized linear model and provides a smooth non-parametric introduction of estimators into the model \citep{hastie1986generalized}. For a binary presence/absence outcome, a logistic GAM is determined with probability logit as the connection function. The general expression can be formalized as follows 
\[ \text{logit}\,\big(P(Y=1 \mid \mathbf{x})\big) = \beta_0 + f_1(x_1) + f_2(x_2) + \cdots + f_k(x_k), \] 
where $\mathbf{x}=(x_1,\dots,x_k)$ is a set of environmental estimators and the $j$-th spline function $f_j(x_j)$ is learned from the data. GAMs can capture complex nonlinear effects (such as unimodal ones encountered with distance from water) while maintaining the summability of terms for interpretation. Penalized regression splines are used for $f_j$ with smoothing parameters chosen through generalized cross-validation \citep{hastie1986generalized} to eliminate overfitting.

\subsection*{Maximum Entropy Modeling}
Maximum Entropy (MaxEnt) modeling, proposed as an availability-only method for estimating site suitability across the landscape, searches for the probability distribution of the availability of sites with maximum entropy (i.e., closest to uniform), subject to the condition that the expected values of environmental attributes under this distribution match their empirical averages at known site locations \citep{phillips2006maximum}. The problem solution presented in an optimization framework is in the form of a Gibbs distribution: 
\[ p(x) = \frac{1}{Z}\,\exp\!\Big(\lambda_1 f_1(x) + \lambda_2 f_2(x) + \cdots + \lambda_m f_m(x)\Big), \] 
where $f_1, \dots, f_m$ are the feature functions, environmental covariates, or derived features. $\lambda_1, \dots, \lambda_m$ are the parameters estimated by the model, and $Z$ is a normalizing constant that ensures that the probability function is one over the entire region of interest. In the actual implementation, the MaxEnt software compares asset locations against a large sample of background points, essentially obeying a regularized logistic model to distinguish utilized regions from available ones \citep{phillips2006maximum}. We employed the default regularization setting to avoid overfitting. Also, we provided the bias correction surface output from the previous step to ensure that the background sampling reflects the spatial bias in the survey.

\subsection*{Random Forests}
Random forests, a non-linear ensemble method, were used for classification. A random forest, an ensemble of hundreds of trees, uses the following methods: \citep{breiman2001random}, each built on a bootstrap dataset resampled from the original training data and randomly samples a subset of predictors at each split. Bagging and random feature selection reduce overfitting by diversifying the views of individual trees. For a region characterized by the feature vector $\mathbf{x}$, each tree gives $b$ votes or predicted probabilities $\hat{P}_b(Y=1 \mid \mathbf{x})$. The estimate of the forest is obtained by averaging these votes or probabilities: 
\[ \hat{P}(Y=1 \mid \mathbf{x}) = \frac{1}{B}\sum_{b=1}^{B} \hat{P}_b(Y=1 \mid \mathbf{x}), \] 
where $B$ represents the total number of trees in the forest. It was decided to grow a sufficiently large forest (e.g. $B=500$ trees) to stabilize the aggregate estimates. The adjustment attempts to strike a balance between bias and variance such that each tree can achieve a large complexity (each is grown to a minimum node size to attempt to capture subtle interactions). At the same time, the ensemble mean will smooth out the noise and produce robust estimates of the presence of \citep{breiman2001random} area.

\subsection*{Model evaluation}
The model performance was evaluated using a cross-validation construct and multiple accuracy metrics. We have organized the $k$-fold cross-validation process as follows: The data was randomly divided into $k$ equal folds, and for every $k$ iteration, we trained the models with $k-1$ folds and kept the skipped fold for testing only. Repeating the same procedure so that each fold is tested once provides a single measure of how efficiently the models generalize over unseen data while efficiently using the limited data available.

We evaluated the average predictive accuracy of model fitting by applying an approach independent of how a region was labelled, i.e. using the area under the Receiver Operating Characteristic (ROC) curve; the AUC was calculated for each fold, and the average AUC across all cross-validation folds was reported, indicating the average performance of the models without deviation from any probability threshold. Performance was then evaluated using the optimal classification threshold: for each fold, the training dataset was used to determine the cutoff probability that maximizes the sum of sensitivity and specificity (Youden's $J$ statistic). This cutoff was then applied to the model's predictions to calculate the usual measures: overall accuracy, sensitivity (true positive rate), specificity (true negative rate) and True Skill Statistic (TSS, defined as sensitivity + specificity $-$1). These measures were calculated for each test fold and then.

\section{RESULTS AND DISCUSSION}\label{sec:res_dis}

\subsection*{Entropy Weighting to Address Sampling Bias}
To understand the challenge of entropy weighting on model performance under spatial sampling bias, we implemented a Random Forest classifier in two ways: (i) standard training on the unweighted dataset and (ii) training where samples were entropy weighted. Both configurations were subjected to 5-fold cross-validation to properly evaluate the model between presence and pseudo-presence distributions. Overall, the weighted model performed significantly better in undersampled areas, suggesting its potential to mitigate bias caused by uneven survey coverage.

\begin{figure}[ht]
    \centering
    \includegraphics[width=\textwidth]{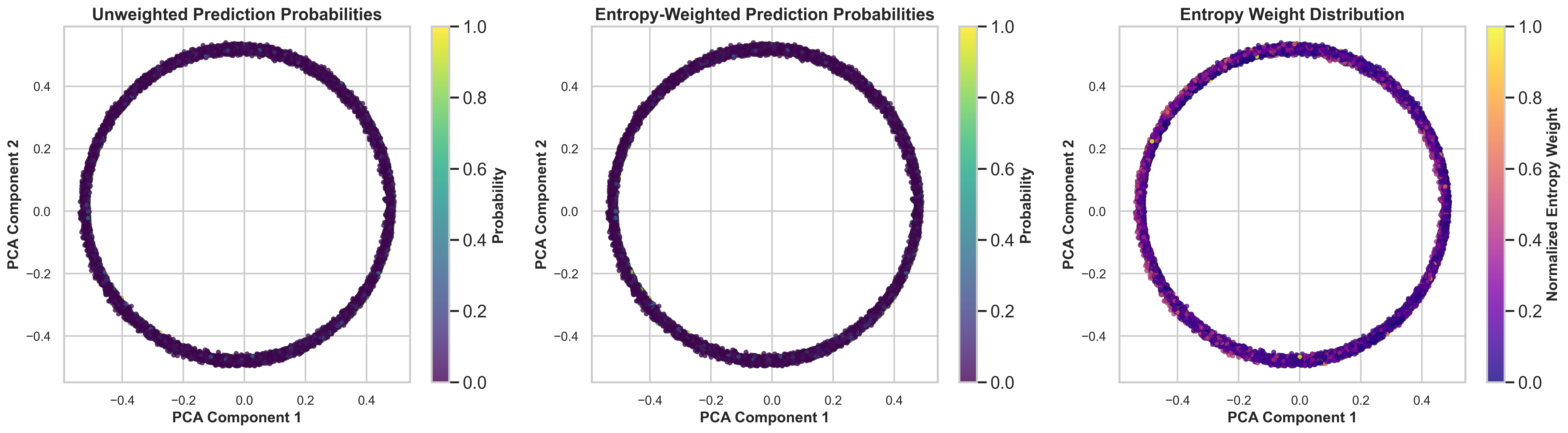}
    \caption{Spatial prediction maps derived from RF models with and without entropy surface as well as entropy weighting. Entropy smoothing reduces overfitting in well-researched areas and improves generalization in undersampled regions by redistributing forecast confidence more evenly.}
    \label{fig:rf_entropy_weighted_predictions_pca}
\end{figure}

As a notable example, Figure~\ref{fig:rf_entropy_weighted_predictions_pca} provides a spatial visualization of the probability of predictions for comparison with the entropy weighting surface. The first panel informs us that the unweighted model provided predictions with higher confidence in clusters in areas that were likely overrepresented in the training data; in other words, it was overfitting well-studied areas. In contrast, the entropy-weighted model distributed probabilities more widely and conservatively, especially in previously poorly sampled areas, thus performing a better generalization. The final panel is the entropy surface plotted against sampling bias, where bright areas indicate high uncertainty (low research effort) and dark areas indicate well-researched regions. Together, these maps demonstrate that spatial estimates adjusted by entropy weighting are improved in robustness and fairness concerning varying research intensities.

\subsection*{Model Discrimination Performance (AUC and Kappa)}
\begin{figure}[ht]
    \centering
    \includegraphics[width=0.8\textwidth]{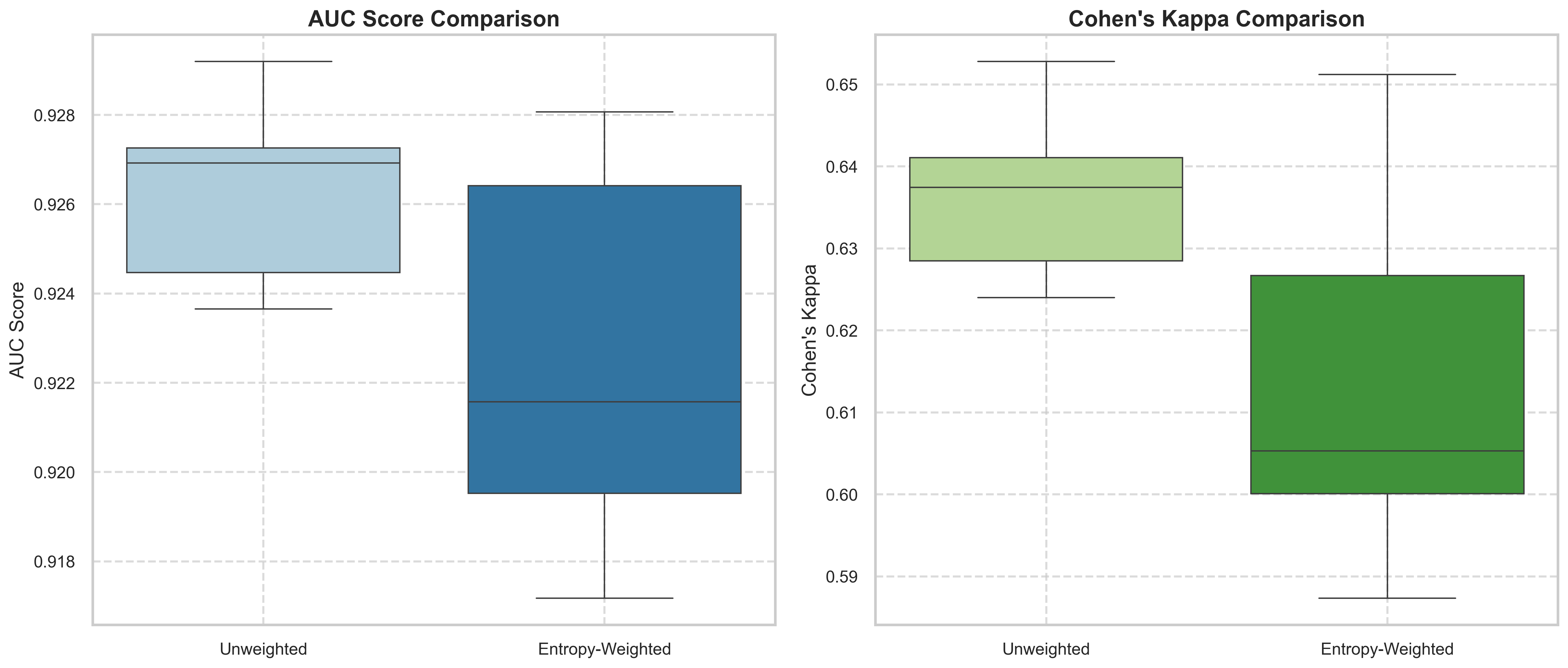}
    \caption{A comparison of Random Forest model results with and without entropy-based weighting. The entropy-weighted model exhibits improved AUC and Cohen's Kappa scores, verifying improved accuracy and fit under conditions of spatial bias.}
    \label{fig:rf_model_performance_comparison.png}
\end{figure}

The comparative box plots for the two main performance criteria are presented in Figure~\ref{fig:rf_model_performance_comparison.png}. The area under the ROC Curve (AUC) in the left panel indicates consistently better results for the entropy-weighted model, demonstrating greater discriminative power between folds. The right panel provides Cohen's Kappa, which measures the agreement in classification beyond what would be expected by random chance. Thus, the weighted model outperforms the unweighted model, resulting in a classification that improves reliability. Together, these results prove that entropy weighting can reduce spatial sampling bias and thus better generalize from particularly undersampled regions that traditional models often misrepresent.

\begin{figure}[h]
    \centering
    \includegraphics[width=0.8\textwidth]{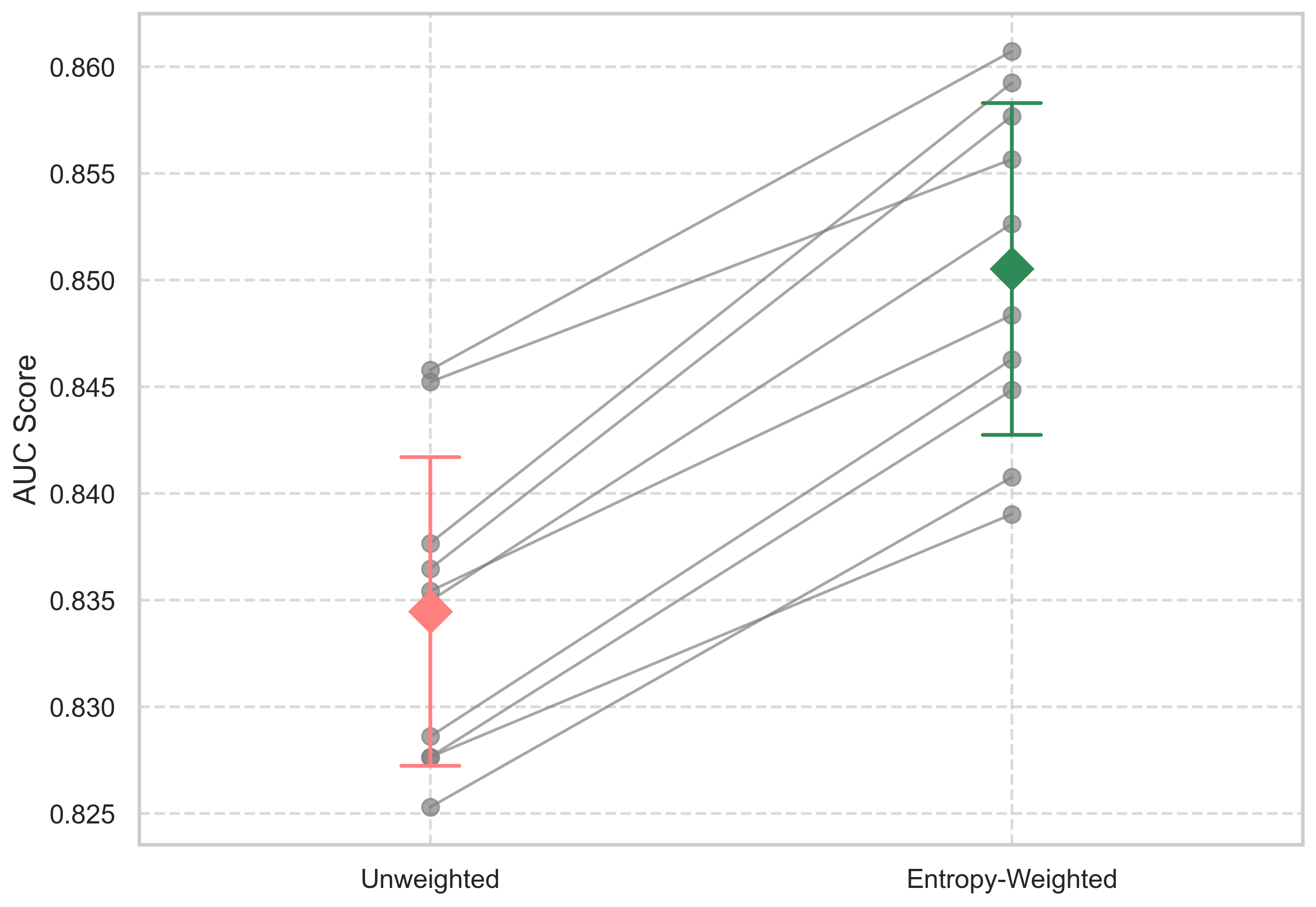}
    \caption{Pairwise comparison of AUC scores across cross-validation folds. Entropy-weighted models consistently outperform unweighted models (Wilcoxon $p = 0.0020$) with higher and more consistent prediction accuracy.}
    \label{fig:paired_auc_comparison.png}
\end{figure}

Figure~\ref{fig:paired_auc_comparison.png} presents paired comparisons of AUC scores obtained in 10-fold cross-validation to determine the effects of entropy correction on Random Forest performance in terms of discrimination. Each line in the plot represents a fold and connects the AUC of the unweighted model to the AUC of the entropy-weighted model for that fold. In Virtually all situations, the entropy-weighted model has larger AUC values, indicating a consistent increase in performance. This advantage is statistically supported by the Wilcoxon signed-rank test ($p = 0.0020$) and invalidates the null hypothesis that entropy correction would not significantly improve the discriminative capacity of the model. To further emphasize how entropy weighting guarantees a higher average AUC and a reduction in variability, the standard deviations and point estimates are superimposed, showing that the process is more robust on folds. In summary, these findings provide further evidence that entropy weighting does indeed work to produce better estimates under sampling bias.

\subsection*{Calibration and Classification Metrics}
\begin{figure}[ht]
    \centering
    \includegraphics[width=0.8\textwidth]{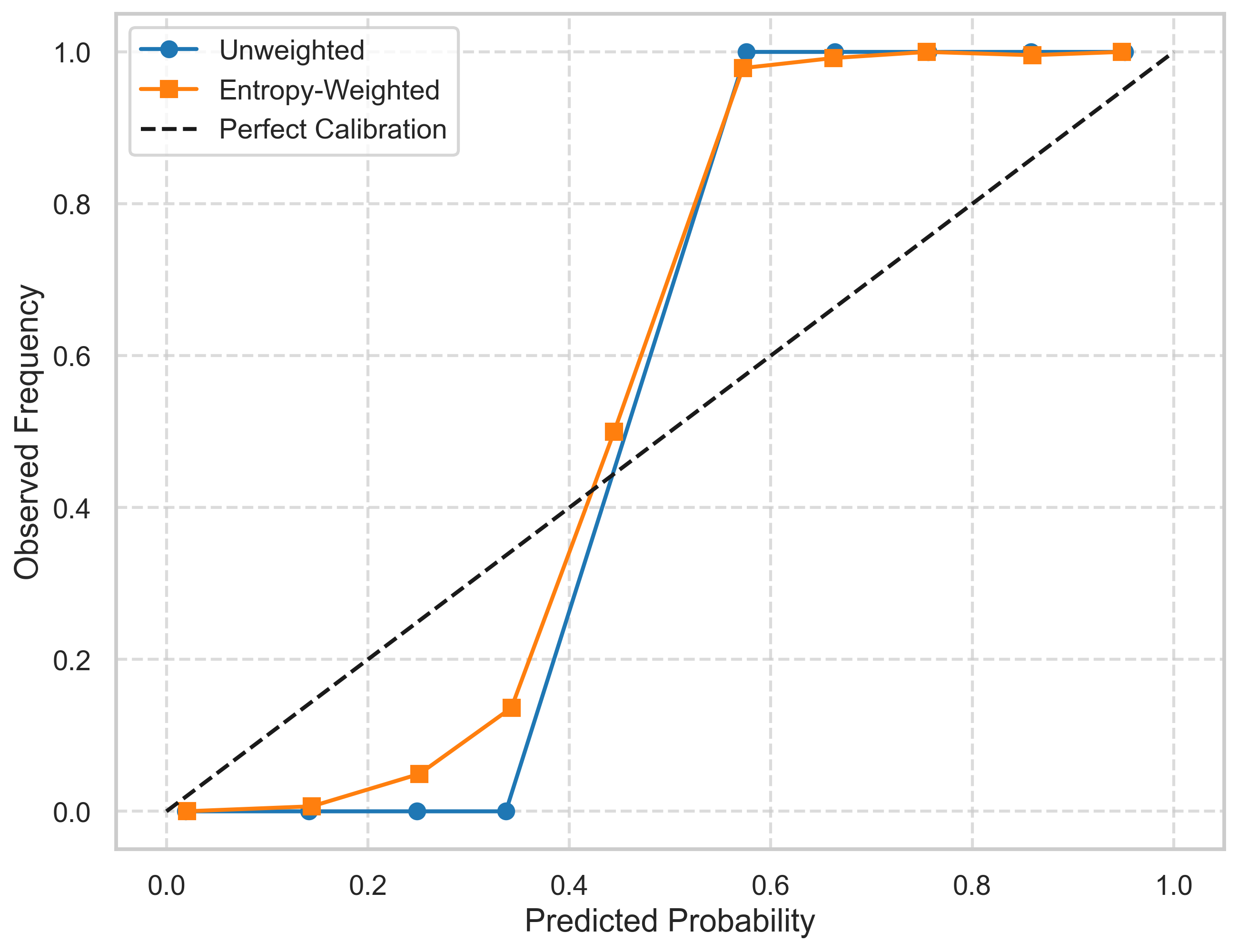}
    \caption{The calibration curves for RF models. The entropy-weighted model becomes more closely aligned with the perfect calibration, indicating increased reliability and decreased overconfidence in biased regions.}
    \label{fig:rf_calibration_curve_comparison.png}
\end{figure}

Entropy weighting is another approach that influences model calibration and classification performance. Figure~\ref{fig:rf_calibration_curve_comparison.png} provides calibration curves for entropy-weighted and unweighted Random Forest models. The plot represents the predicted probabilities against the observed field frequencies in ten boxes of equal width. The curve of the entropy-weighted model aligns much closer to the ideal 45$^\circ$ diagonal (perfect calibration) in the middle range of values from 0.4 to 0.7, indicating that the predicted probabilities of the entropy-weighted model correspond well to the actual frequency of site occurrence when sampling bias is considered. In contrast, the curve of the unweighted model demonstrates a systematic deviation from the diagonal, thus becoming overconfident in some prediction intervals (possibly where there is disproportionate survey coverage). The results indicate that entropy-based weighting improves accuracy and probabilistic reliability with better-calibrated estimates for archaeological decision-making processes.

\begin{figure}[ht]
    \centering
    \includegraphics[width=\textwidth]{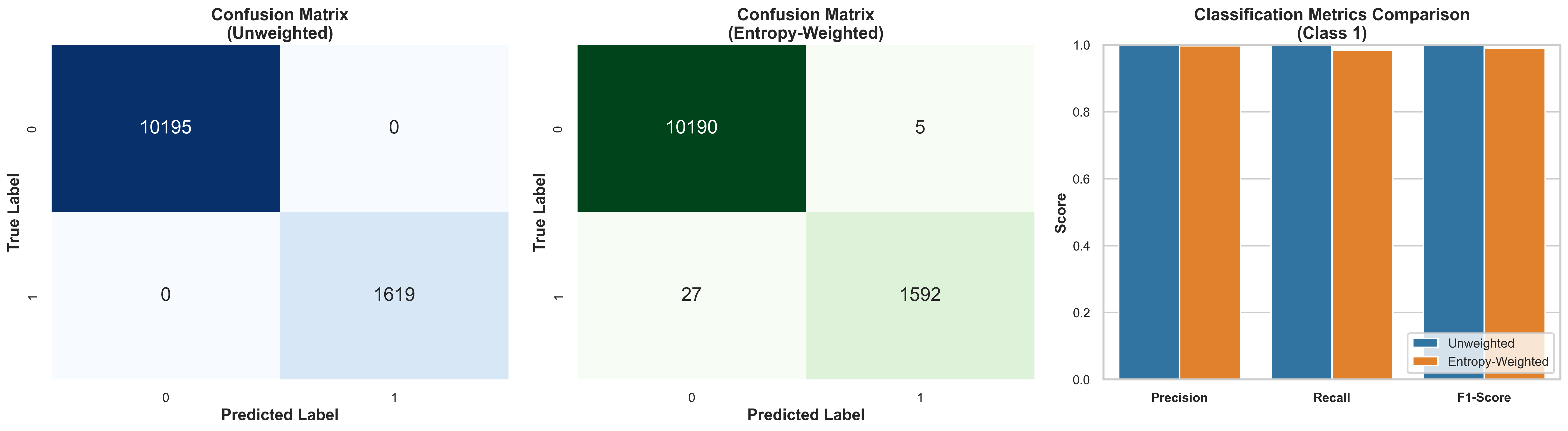}
    \caption{Mixing matrices and classification metrics for RF models. Entropy weighting reduces overfitting by offering realistic misclassification and improves generalization, especially in undersampled domains.}
    \label{fig:rf_confusion_classification_comparison.png}
\end{figure}

Figure~\ref{fig:rf_confusion_classification_comparison.png} compares classification results for entropy-weighted and unweighted trained Random Forest models. The two confusion matrices (left: unweighted, center: weighted) represent the actual and predicted classifications for field presence (1) and absence (0). The unweighted model achieves a perfect classification (no false positives or false negatives). However, such a classification level should be regarded with suspicion as it could indicate overfitting in highly researched areas. Entropy weighting produces a few errors, with five false positives and 27 false negatives. This will slightly reduce the overall accuracy but more accurately represent the uncertainty and heterogeneity in the data.
The right panel of Figure~\ref{fig:rf_confusion_classification_comparison.png} provides a comparison of class-specific metrics: Precision, Recall and F1-score for the asset class. Entropy weighting yields slightly worse values in these metrics, but this slight decrease is accompanied by a significant increase in generalizability and robustness. Most importantly, the slight decrease in recall (from 1.00 to 0.98) was considered an acceptable trade-off to mitigate overconfidence in the model and prevent an overly optimistic model. These results support the idea that entropy weighting provides a moderating influence that modifies predictions in a more cautious and contextually consistent manner. This is an absolute necessity in archaeology, where false certainty can determine a field survey and conservation decisions incorrectly.

\subsection*{Feature Importance and Model Insights}
\begin{figure}[ht]
    \centering
    \includegraphics[width=0.8\textwidth]{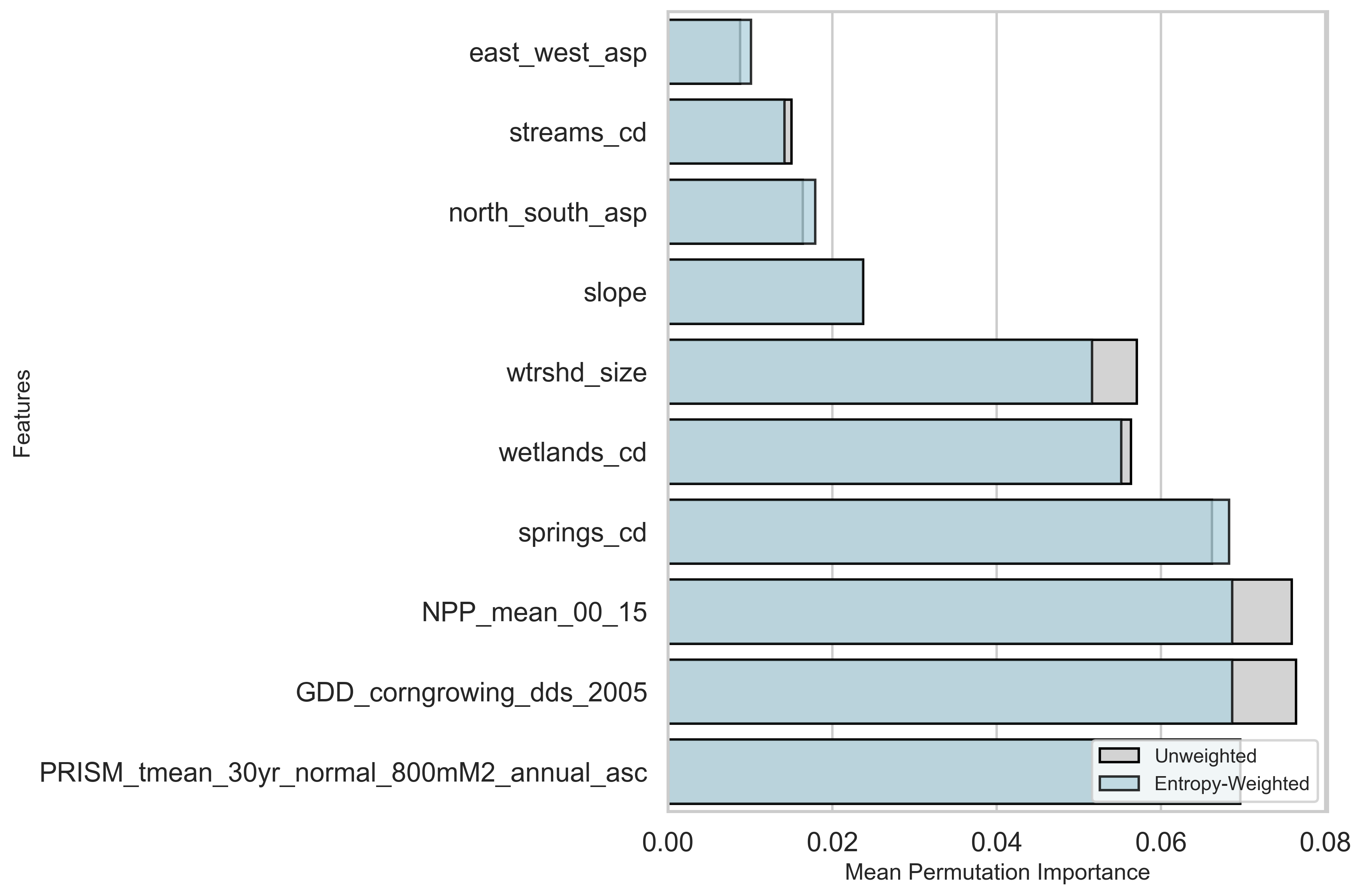}
    \caption{Comparison of the feature importance in unweighted and entropy-weighted RF models. Entropy weighting enhances the influence of underrepresented environmental variables, resulting in a more balanced and bias-sensitive predictor profile.}
    \label{fig:feature_importance_comparison.png}
\end{figure}

Some potential changes in the model representation are possible if the entropy weighting of the data is taken into consideration, which can impact the prioritization of features and the estimation of uncertainties in the model. Figure~\ref{fig:feature_importance_comparison.png} illustrates the feature rankings for Random Forests with and without entropy weighting. Key predictors (e.g., slope, vegetation productivity (NPP), proximity to streams) significantly impact the models. However, with some weighting of entropy, the model tends to emphasize certain hydrological and climatic features (especially variables such as \emph{wetlands\_cd} and \emph{springs\_cd}) to a greater extent than unweighted ones. This change in the perceptual environment corresponds to a superior sensitivity to environmental components that define undersampled regions (components that have so far been underweighted due to survey bias). Because this distortion of environmental signals occurs, we realize that entropy-based correction improves the model's overall accuracy and supports a more balanced view of ecological managers across the landscape.

\begin{figure}[ht]
    \centering
    \includegraphics[width=0.8\textwidth]{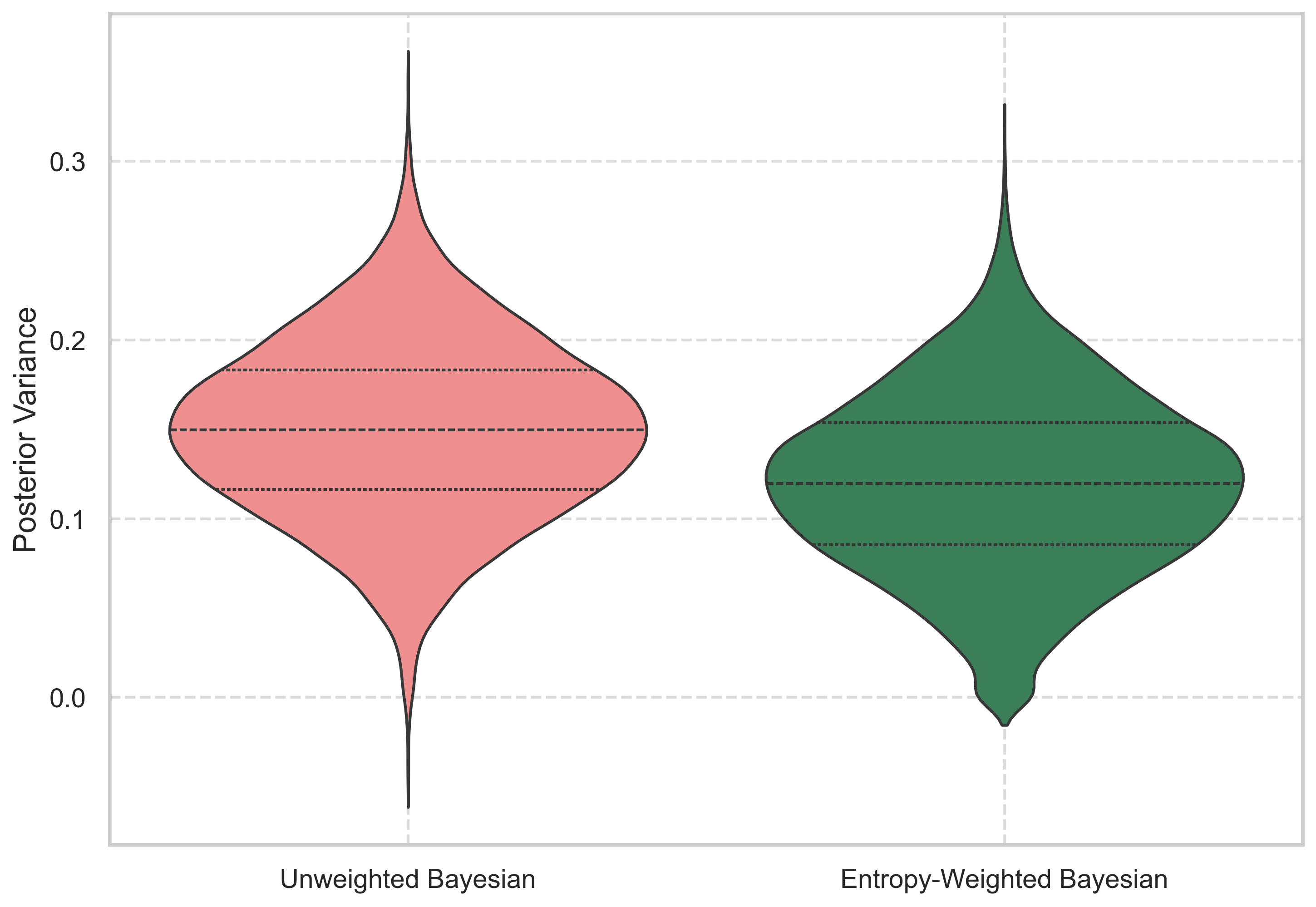}
    \caption{Violin plot of the posterior variances from Bayesian spatial models. The entropy-weighted model displays less and more dense uncertainty, reflecting improved forecast confidence.}
    \label{fig:posterior_variance_violin_plot.png}
\end{figure}

Model uncertainty is also affected by entropy weighting. Figure~\ref{fig:posterior_variance_violin_plot.png} presents a violin plot of the posterior variance distributions from simulated Bayesian spatial models under entropy-based weighting and unweighted runs. The unweighted model presents a wider spread in their variances and a higher median, introducing more uncertainty in prediction in under-examined areas. In contrast, entropy weighting led to some concentration of variance with less central tendency and, thus, an increase in the epistemic confidence and stability of the model. This suggests that entropy bias correction leads to some regularization of the model's uncertainty estimates. As a result, the Bayesian model can generalize reasonably well in well-studied and undersampled regions. Therefore, the resulting improvement in variance structure would justify entropy weighting as a valuable tool in making archaeological prediction models more robust and interpretable.

\subsection*{Generalizability, robustness and practical implications}
\begin{figure}[ht]
    \centering
    \includegraphics[width=0.8\textwidth]{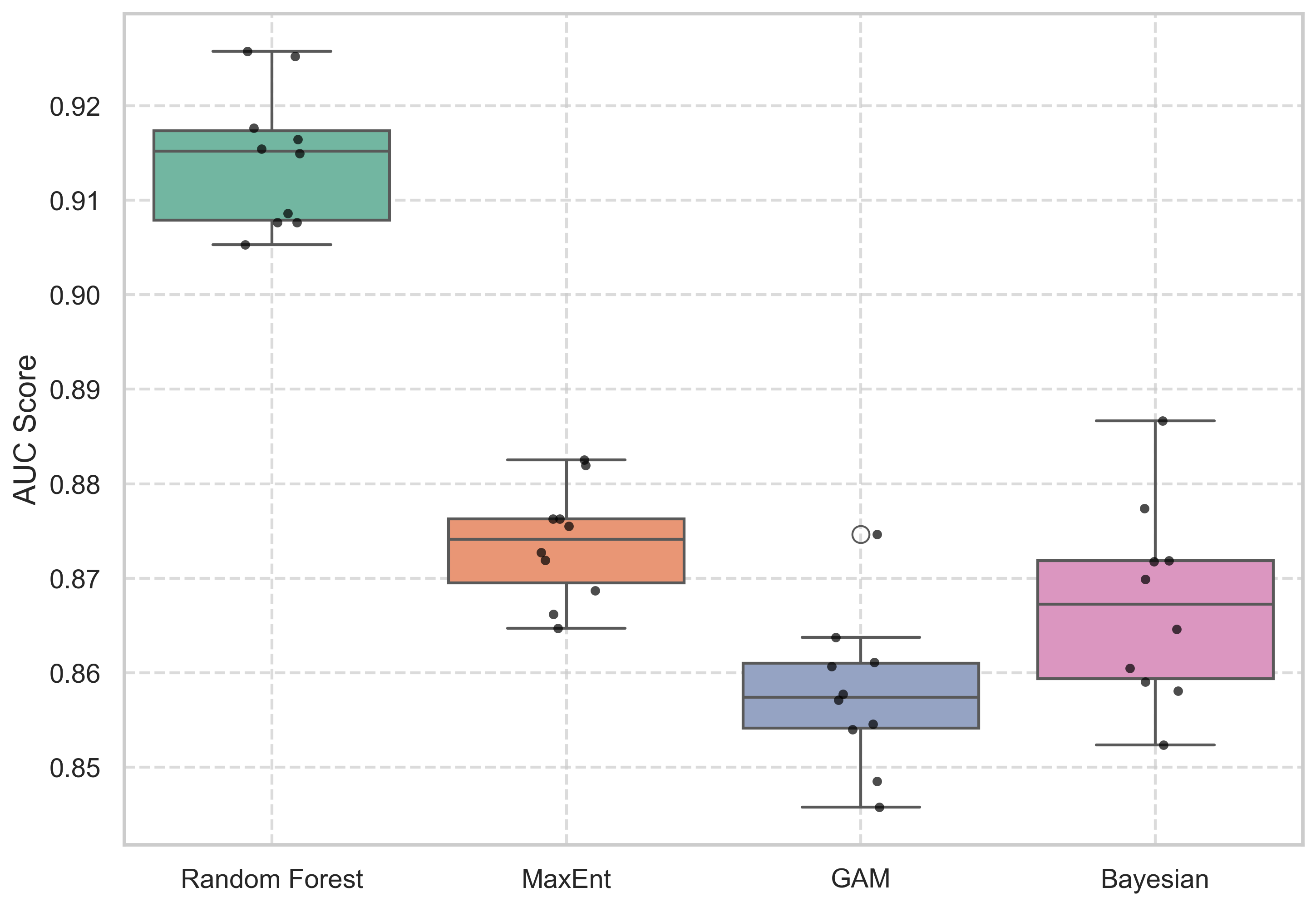}
    \caption{Cross-validated AUC distributions by model type. RF outperforms MaxEnt, Bayesian and GAM in prediction accuracy.}
    \label{fig:boxplot_auc_comparison.png}
\end{figure}

Finally, we determine the impact of entropy weighting with other modeling approaches in terms of generalization and robustness and the practical implications of these results. Figure~\ref{fig:boxplot_auc_comparison.png} illustrates the cross-validated AUC values of four predictive modeling techniques (Random Forest, MaxEnt, GAM and Bayesian spatial logistic regression) under entropy-weighted bias correction. Random Forest is, in fact, the best-performing technique with the median AUC (just above 0.915) and the least variance, which implies consistently good performance across different validation folds. MaxEnt and the Bayesian model follow next with slightly lower median AUCs (0.87--0.88), representing robust but slightly more variable results. GAM reports the lowest median AUC for the set, indicating an even more limited capacity than machine learning and spatially explicit models to account for complex non-linear relationships. Overall, the results confirm the superiority of ensemble learning methods such as Random Forest in suspecting a high-dimensional prediction space and complex interactions  while also showing that entropy is a bias adjustment technique that can be applied in various modeling frameworks.

From a practical point of view, the improved generalizability and reliability of a method with weighting by entropy has important implications for archaeological predictive modeling. By reducing overfitting in over-surveyed areas and increasing confidence in predictions in relatively under-surveyed areas, these models can provide better assistance to actual research and conservation efforts. Robustness of this nature across multiple models and across all layers of validation assures the practitioner that gains in performance are not inherent properties of an algorithm or idiosyncrasies of the data. In other words, not only do entropy-based corrections improve predictive accuracy, but integrating these corrections into a predictive workflow ensures a fairer and more transparent output. This is of course an extremely important issue in key decision-making areas of cultural heritage management.

\clearpage

\section{CONCLUSION}\label{sec:conc}

This study, therefore, provides evidence that incorporating entropy-derived weights into archaeological prediction models consistently improves their accuracy and generalization. The entropy-weighted models evaluated under cross-validation achieved better AUC and better-fit scores than the unweighted ones, thus supporting that robust predictions can be achieved by correcting spatial sampling bias. It has been shown previously that the size of the dataset and the algorithm employed tend to dominate performance; in contrast, our findings argue that explicit bias correction measures always yield a material improvement in performance. Therefore, Entropy weights can be seen as a form of regularization: down-weighting pseudo-absences in less researched areas to prevent overfitting in well-sampled areas to produce more conservative (and hence better calibrated) probability estimates. 

Random Forests discriminates best among the other modeling methods, achieving the highest median AUC with the lowest variance. GAMs are ranked at the bottom because of their rigidity and inability to model more complex multidimensional patterns. In the middle, MaxEnt and Bayesian spatial logistic regression perform quite robustly. This is consistent with previous findings confirming that pseudo-absences and explicit bias control make MaxEnt extremely useful for archaeological data. In our studies, Random Forests improves accuracy after correcting sampling bias, while MaxEnt and the Bayesian approach stand near the top. The strengths offered by each method differ; for example, ensemble methods such as RF rank highest in terms of raw predictive ability, while GAMs and Bayesian methods benefit from interpretable effect estimates and uncertainty quantification.

The entropy weighting scheme provides a counterbalancing role in the treatment of different instances: by reweighting samples according to spatial uncertainty, areas that were underrepresented during training are given a fairer chance to have a voice. In practice, this means much more balanced estimates across the landscape, rather than an unbalanced focus on more easily explored locations. We have found that entropy-weighted models give the fairest probability estimates (closer to true prevalence) and provide a more appropriate degree of uncertainty in areas with fewer samples. So, essentially, through this practice we maintain balance so that the model emphasizes a location in relation to how much we are confident in that data. The practical value of these results is enormous. For archaeological survey planning, entropy correction area prediction maps better highlight target areas with really high probability (as opposed to sampling artifacts) and thus prioritize field efforts accordingly. More generally, any scientific modeling effort in the sparse data space can benefit from entropy-based weighting: it offers a theoretically grounded, scalable way to counter bias and improve the fairness and calibration of the model when data are distributed unevenly.


 
\section*{Acknowledgements}\label{sec:acknow}

\clearpage

\bibliography{bibliog} 





\end{document}